\documentclass[final,5p,twocolumn]{elsarticle}
\journal{Physics Letters B}

\usepackage{graphicx,color,amsmath,amssymb}
\usepackage{bm}
\usepackage[colorlinks=true, pdfstartview=FitV, linkcolor=red, citecolor=blue, urlcolor=blue]{hyperref}

\begin{document}

\begin{frontmatter}

\title{ The rapidity dependence of the average transverse momentum in 
p+Pb collisions at the LHC: the Color Glass Condensate versus hydrodynamics. }

\author[pb]{Piotr Bo\.zek}
\address[pb]{AGH University of Science and Technology, 
Faculty of Physics and Applied Computer Science, PL-30-059 Krak\'ow, Poland
\\ 
Institute of Nuclear Physics PAN, PL-31342 Krak\'ow, Poland}
\ead{piotr.bozek@ifj.edu.pl}

\author[ab]{Adam Bzdak}
\address[ab]{RIKEN BNL Research Center, Brookhaven National Laboratory, Upton, NY 11973, USA}
\ead{abzdak@bnl.gov}

\author[vs]{Vladimir Skokov}
\address[vs]{Department of Physics, Western Michigan University, Kalamazoo, MI 49008, USA}
\ead{vskokov@quark.phy.bnl.gov}

\begin{abstract}
We show that in proton-lead (p+Pb) collisions at the LHC,
the Color Glass Condensate (CGC) and hydrodynamics lead to 
qualitatively different behavior of the average transverse momentum, 
$\langle p_{\perp} \rangle$, with the particles rapidity.
In hydrodynamics, the $\langle p_{\perp} \rangle$ decreases as one
goes from zero rapidity, $y=0$, to the proton fragmentation region since
the number of particles decreases.
In contrast, in the CGC the saturation momentum increases as 
one goes from $y=0$ to the proton fragmentation region, and so the $\langle p_{\perp} \rangle$ 
increases. At the LHC, the difference between the two models may be large enough
to be tested experimentally.
\end{abstract}

\end{frontmatter}

% $\langle p_{\perp} \rangle$

\section{Introduction}

Recent experiments with proton-lead (p+Pb) collisions at the LHC  
on 2- and 4-particle correlations 
\cite{CMS:2012qk,Abelev:2012ola,Aad:2012gla,Aad:2013fja,Chatrchyan:2013nka,ABELEV:2013wsa} 
give raise to various theoretical interpretations. The two-dimensional correlation
functions in relative pseudorapidity  and relative azimuthal angle  
demonstrate 
the ridge--like structures, elongated in pseudorapidity,
with enhanced emission of particle pairs in same $\Delta \phi \simeq 0$ and 
away-side $\Delta  \phi \simeq \pi$ directions. 
The Color Glass Condensate (CGC) approach leads to the long-range correlations in 
rapidity \cite{Dusling:2009ni} with the same-side 
structure coming from the interference diagrams enhanced in the saturation 
regime \cite{Dusling:2013oia,Dusling:2012iga}. The measured 
elliptic and triangular harmonic coefficients of azimuthal distributions
can also be explained by the hydrodynamic expansion of the dense small fireball, see Refs.
\cite{Bozek:2011if,Bozek:2013uha,Bzdak:2013zma,Qin:2013bha,Werner:2013ipa,Bozek:2013ska}. 
A recently proposed measurement of the femtoscopy 
radii in p+Pb interactions at different centralities could disentangle between the two scenarios 
\cite{Bozek:2013df,Bzdak:2013zma}, as the collective expansion leads generally to a larger size of the system.
Another observable sensitive to the collective expansion is  the  average transverse 
momentum $\langle p_\perp \rangle$ of the emitted particles
\cite{ALICE:2012mj,Chatrchyan:2013eya,Abelev:2013bla}. The average momentum is larger in proton-proton than in 
p+Pb collisions for events of the same multiplicity.  
This fact can be understood in the model that treats a p+Pb collision as a superposition of independent p+p interactions. For p+Pb collisions, the model also gives the results which are below the experimental data \cite{Bzdak:2013lva}.
This leaves room for an additional collective push, which is naturally present in hydrodynamics. 
Also the increase of  $\langle p_\perp \rangle$
with the particle mass observed in p+Pb interactions can be quantitatively understood in hydrodynamics
\cite{Bozek:2013ska,Chatrchyan:2013eya}. On the other hand, the mass hierarchy 
of the transverse momentum at central rapidity 
may appear due to the color reconnection or the geometrical scaling discussed in 
Refs.~\cite{Ortiz:2013yxa,McLerran:2013oju}.  
To resolve this ambiguity in p+Pb collisions, the measurement of the number of charged particles 
at central rapidity as a function of 
the number of participants was proposed in Ref.~\cite{Bzdak:2013zla}.

In this letter, we propose to study the rapidity, $y$, dependence of the average transverse momentum of 
charged particles. In the CGC the average transverse momentum is determined 
by the nucleus saturation momentum. The evolution of the saturation momentum with rapidity towards 
the proton direction yields a growth of the average
transverse momentum, quite in opposite to what is expected from a collective expansion.
Namely, the hydrodynamic model predicts a decrease of the average transverse 
momentum when going from midrapidity, $y=0$, to the proton side, owing to a decreasing number 
of produced particles.

\section{Rapidity dependence of transverse momentum}

In the CGC the dependence of the average transverse
momentum on  rapidity can be deduced from quite general
arguments. First of all, 
the relation between the average transverse momentum of final
particles, $\left\langle p_{\perp }\right\rangle $, and the average
transverse momentum of produced gluons, 
$\left\langle k_{\perp }\right\rangle$, is 
\begin{equation}
\left\langle p_{\perp }\right\rangle =\frac{\int d^{2}p_{\perp}  p_\perp \int_0^1 dz\frac{D(z)%
}{z^{2}}f_{g}(\frac{p_{\perp }}{z})}{\int d^{2}p_{\perp}\int_0^1 dz\frac{D(z)}{%
z^{2}}f_{g}(\frac{p_{\perp }}{z})}=\left\langle z\right\rangle \left\langle
k_{\perp }\right\rangle ,
\label{ptok}
\end{equation}%
where $f_{g}(\frac{p_{\perp }}{z}=k_{\perp })$ is the 
gluon distribution function, and $D(z)$ is the gluon fragmentation function. We defined the first 
moment of the gluon fragmentation function $\langle z \rangle$ as 
\begin{equation}
\left\langle z\right\rangle =\frac{\int_{0}^{1}dzD(z)z}{\int_{0}^{1}dzD(z)}.
\end{equation}
Deriving Eq.~\eqref{ptok} we assumed that the gluon fragmentation function is independent of 
the transverse momentum. This assumption may not be justified for very soft gluons.\footnote{The average
transverse momentum of produced pions in high multiplicity p+Pb collisions is approximately $0.6$ GeV,
thus $\left\langle k_{\perp }\right\rangle$ is expected to be around a few GeV.}  

The information available about the gluon fragmentation function is rather limited. 
Therefore it is important to construct an observable for which $\langle z \rangle$
cancels out. In this letter, we adopt the ratio of the transverse momentum at a given
rapidity $y$ to the value at $y=0$: 
\begin{equation}
\frac{\left\langle p_{\perp }\right\rangle _{y}}{\left\langle p_{\perp
}\right\rangle _{y=0}}=\frac{\left\langle k_{\perp }\right\rangle _{y}}{%
\left\langle k_{\perp }\right\rangle _{y=0}}.  \label{pt-ratio}
\end{equation}

The gluon distribution function can be obtained within the $k_{\perp}$-factorization formalism, 
according to which the cross-section
for inclusive gluon production reads~\cite{Kovchegov:2001sc}: 
\begin{equation}
	\frac{d \sigma^{p+A\to g}}{d^2 k_\perp dy } = \frac{2 \alpha_s}{C_F} \frac{1}{k^2_\perp} \int d^2q_\perp \phi_p(q_\perp^2) 
	\phi_A\left( (\vec{k}_\perp-\vec{q}_\perp)^2 \right), 
\label{ktfac}
\end{equation}
where $\phi_{p, A}$ are the unintegrated gluon distribution (UGD) functions for the proton and 
the nucleus, respectively, and the Casimir operator in the fundamental representation of SU(3) is given by $C_F = 4/3$.
Here and in what follows, to lighten the notation  we suppress dependence of the UGD and saturation momentum on $x$. 
The gluon distribution is then 
\begin{equation}
f_g(k_\perp) = \frac{dN}{d^2 k_\perp dy} = \frac{1}{\sigma_{\rm inel}} \frac{d \sigma^{p+A\to g}}{d^2 k_\perp dy }. 
\label{f_g}
\end{equation}
Using the McLerran-Venugopalan model for the classical gluon distribution function one gets (see Ref. \cite{Dumitru:2001ux} for details)~\footnote{%
This relation is valid in the regime $Q_{p}<k_{\perp }<Q_{A}$.}%
\begin{equation}
\left\langle k_{\perp }\right\rangle \approx 2Q_{A}\frac{\ln (\frac{Q_{A}}{%
Q_{p}})-1+\frac{Q_{p}}{Q_{A}}}{\ln ^{2}(\frac{Q_{A}}{Q_{p}})}.
\label{kt-log}
\end{equation}%
Neglecting logartithmic corrections we have for the gluon distribtuation function
\begin{equation}
\frac{f_{g}(k_{\perp })}{S_{\perp }}\propto \left\{ 
\begin{array}{cc}
1, & k_{\perp }<Q_{p}, \\ 
\frac{Q_{p}^{2}}{k_{\perp }^{2}}, & Q_{p}<k_{\perp }<Q_{A}, \\ 
\frac{Q_{p}^{2}Q_{A}^{2}}{k_{\perp }^{4}}, & k_{\perp }>Q_{A}.%
\end{array}%
\right. 
\label{fg}
\end{equation}
This formula captures the general features of the CGC description of p+Pb
collisions, see discussions in Refs.~\cite{Dumitru:2001ux,Kharzeev:2004if}. 
In this framework the system is characterized by two different
saturation scales: $Q_{p}$, the saturation momentum of the proton, and $Q_{A}$
the saturation momentum of the nucleus. In our discussion we assume that 
$Q_{A}\gg Q_{p}$, which seems to be justified for central p+Pb
collisions. Performing straightforward integrations we obtain
\begin{equation}
\left\langle k_{\perp }\right\rangle =\frac{2Q_{A}-\frac{2}{3}Q_{p}}{1+\ln (%
\frac{Q_{A}}{Q_{p}})}\approx \frac{2Q_{A}}{1+\ln (\frac{Q_{A}}{Q_{p}})}.
\label{kt}
\end{equation}%
As expected the average transverse momentum of gluons is roughly
proportional to the saturation momentum of the nucleus. Taking into account 
certain logarithmic corrections to Eq. (\ref{fg}), one obtains a
more accurate expression Eq.~\eqref{kt-log}.

The rapidity dependence of 
the average transverse momentum follows from the standard relations (see, e.g., Ref.~\cite{Khoze:2004hx})
\begin{equation}
Q_{A}^{2}\sim Q_{0}^{2}N_{\text{\textrm{part}}}^{\mathrm{Pb}}e^{\lambda y},
\label{QAy}
\end{equation}%
\begin{equation}
Q_{p}^{2}\sim Q_{0}^{2}e^{-\lambda y}.
\end{equation}
In this article we choose $\lambda \approx 0.2$, following Ref.~\cite{Dumitru:2011wq}.
It is worth noticing that our results for $%
\left\langle p_{\perp }\right\rangle _{y}/\left\langle p_{\perp
}\right\rangle _{y=0}$ are insensitive to the value of $Q_{0}$.

\begin{figure}[t]
\begin{center}
\includegraphics[scale=0.45]{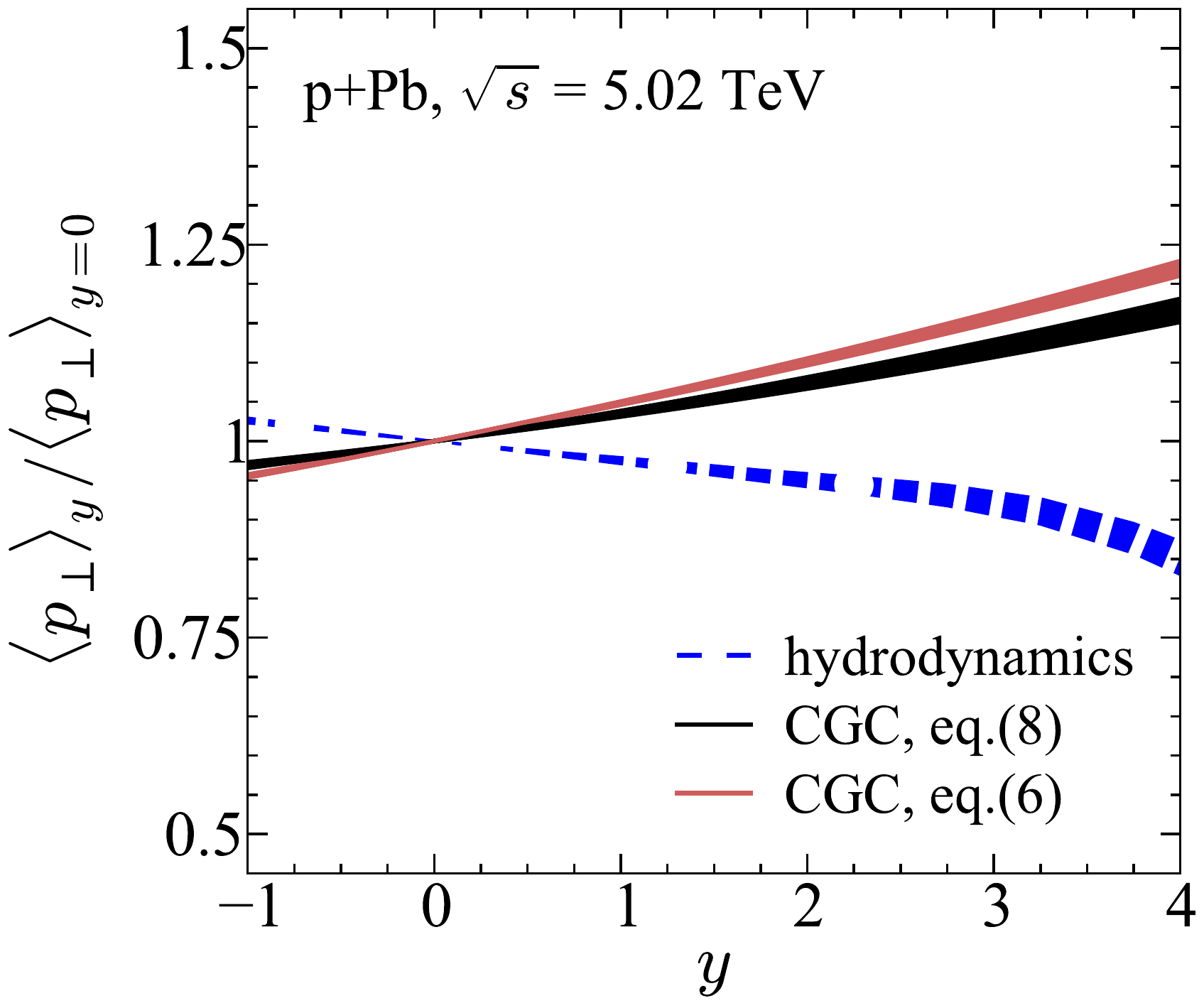}
\end{center}
\caption{The average transverse momentum of produced particles as a function
of rapidity,  divided by the average transverse momentum at $y=0$. The
CGC results for different values of $N_{\mathrm{part}}^{\mathrm{Pb}}$
(black and red bands corresponding respectively to Eqs. (\ref{kt}, \ref{kt-log})) differ
qualitatively from those obtained in the hydrodynamical framework (the dashed band covers three 
centralities $0-3$, $5-15$, and $40-60$\%).}
\label{fig:1}
\end{figure}

Substituting above relations to Eqs. (\ref{kt}, \ref{kt-log}) we obtain the
results presented in Fig. \ref{fig:1}. The black band corresponds to
calculations based on Eq. (\ref{kt}) with several values of $N_{\text{part}}$, ranging from $10$ to $25$. 
The red band is based on Eq. (\ref{kt-log}). As expected the $\langle p_{\perp} \rangle$ in the CGC is increasing
when going from $y=0$ towards the proton fragmentation region owing to increasing $Q_A$, see Eq. (\ref{QAy}).
It is worth mentioning that 
$\left\langle p_{\perp }\right\rangle _{y}/\left\langle p_{\perp
}\right\rangle _{y=0}$ very weakly depends on $N_{\rm part}^{\rm Pb}$ since 
$\langle p_{\perp} \rangle \sim Q_A$ and the number of participants cancels in the ratio.
We emphasize that our results are not sensitive to the specific form of $D(z)$, see Eq. (\ref{pt-ratio}).

In hydrodynamics the dependence of $\left\langle p_{\perp }\right\rangle
_{y}/\left\langle p_{\perp }\right\rangle _{y=0}$ on rapidity is expected to
be quite opposite. 
The average transverse momentum of particles emitted in the  hydrodynamic model is composed of two contributions
the thermal motion at the freeze-out and the collective velocity acquired during the expansion.
Unlike in heavy-ion  collisions, in p+Pb interactions the matter density  depends strongly on rapidity.
Experimental results  show a larger  multiplicity on the lead side than on the proton side \cite{ALICE:2012xs}, 
and the asymmetry increases with centrality of a collision \cite{ATLAS-CONF-2013-096}. 
In  hydrodynamics the collective 
flow velocity results from the action of pressure gradients in the fireball 
\cite{Ollitrault:1991xx}. The initial energy deposition in the fireball should increase as function of
space-time rapidity when going to the lead side in order to match the observed asymmetry of 
charged particle density in pseudorapidity.  

\begin{figure}[t]
\begin{center}
\includegraphics[scale=0.45]{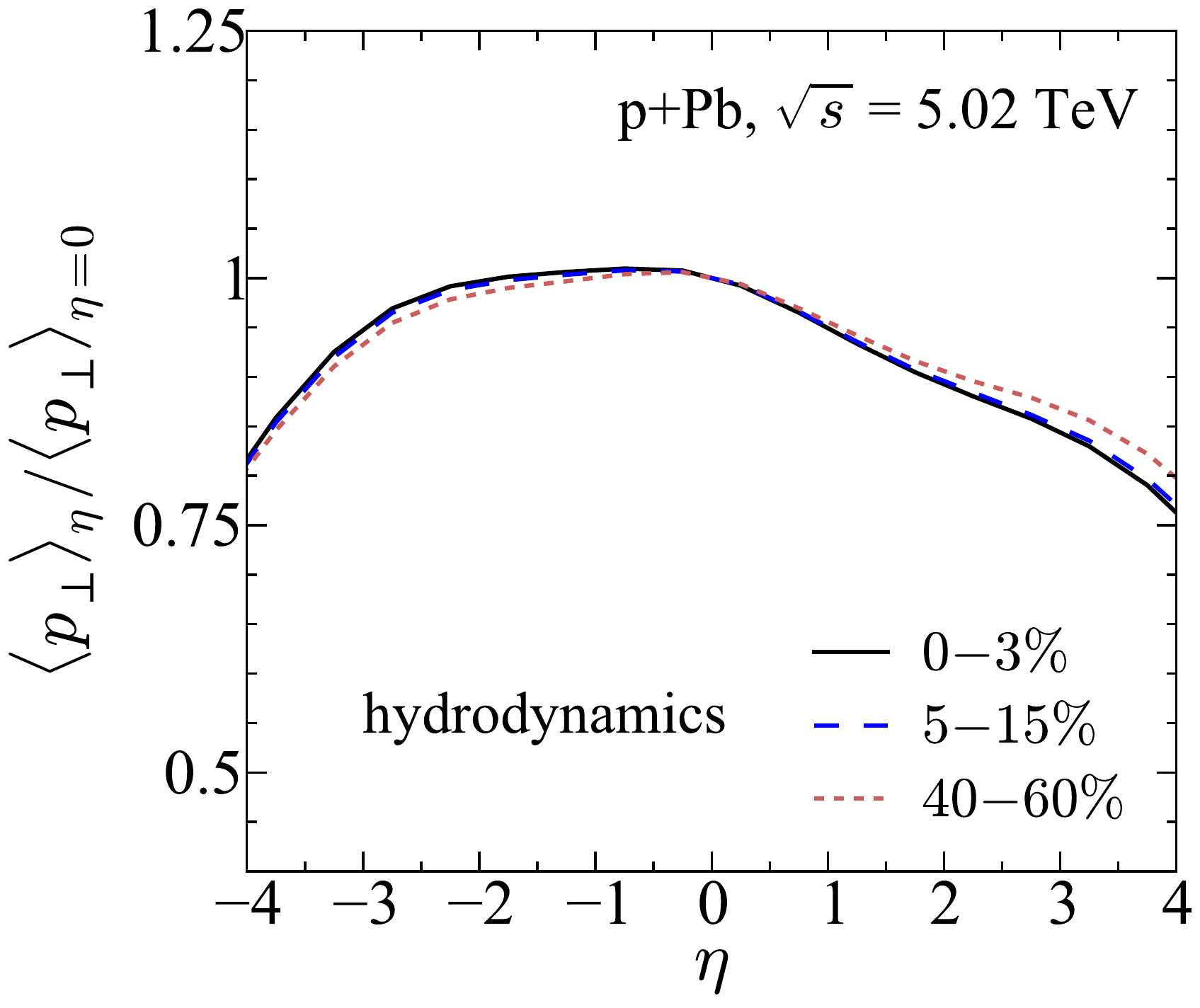}
\end{center}
\caption{The average transverse momentum of produced particles as a function
of pseudorapidity, $\eta$, divided by the average transverse momentum at $\eta=0$, obtained from 
the hydrodynamical calculations at three centralities.}
\label{fig:2}
\end{figure}

In Figures \ref{fig:1} and \ref{fig:2} we present the results obtained from state-of-the-art 
$3+1$ dimensional event-by-event hydrodynamic simulations \cite{Bozek:2011ua}. In this calculation, 
the asymmetry of the initial density of the fireball is imposed 
following the experimental observations in deuteron-gold collisions at $\sqrt{s_{NN}}=200$~GeV 
\cite{Back:2003ns,Bialas:2004su,Bialas:2007eg}. The initial entropy profile is determined by the positions of the 
participant nucleons obtained from the Glauber Monte Carlo model. The entropy deposited at the transverse
position $x,y$ and space-time  rapidity $\eta_\parallel$ by 
a participant located at the position $x_i,y_i$ is
\begin{equation}
s_i(x,y,\eta_\parallel)=f_\pm(\eta_\parallel) \exp \left(-\frac{(x-x_i)^2+(y-y_i)^2}{2 \sigma_w^2}\right), 
\label{eq:sdens} 
\end{equation}
where $\sigma_w=0.4$ fm.
The profiles  $f_{\pm}(\eta_\parallel)$ are 
of the form 
\begin{equation}
f_{\pm}(\eta_\parallel)=\left(1\pm \frac{\eta_\parallel}{y_{\rm beam}}\right)f(\eta_\parallel) \ 
\label{eq:asy}
\end{equation}
with the longitudinal profile
\begin{equation}
f(\eta_\parallel)=\exp\left(-\frac{(|\eta_\parallel|-\eta_0)^2}
{2\sigma_\eta^2}\theta(|\eta_\parallel|-\eta_0)
\right) \ ,
\label{eq:etaprofile}
\end{equation}
where 
$\sigma_\eta=1.4$, $\eta_0=2.4$, and $y_{\rm beam}=8.5$ is the beam rapidity.  The total entropy is 
the sum of the contribution
of the incoming proton and $N_{\rm part}^{\rm Pb}$ nucleons from the lead nucleus, defined with 
the signs ``$+$'' and ``$-$'' respectively in Eq. \eqref{eq:asy}. With the increasing number 
of participants the asymmetry of  the fireball increases, 
yielding the charged particle pseudorapidity distributions in semi-quantitative agreement with
experiment \cite{ATLAS-CONF-2013-096}. 

The  parameters for the hydrodynamic calculation  are chosen as in Ref.  \cite{Bozek:2013ska},
so that it  reproduces reasonably the transverse momentum of 
identified particles in central and semi-central collisions as well as the elliptic and the triangular 
flow in the most central collisions. As seen in Figs. \ref{fig:1} and \ref{fig:2} the transverse 
momentum for various centralities decreases 
when going from $y=0$ to the proton side. The precise form of the charged particle density and of the average
transverse momentum obtained from the hydrodynamic model depends on the parameters of the initial profile 
in Eq. \eqref{eq:etaprofile} and on the details of the Glauber model used \cite{Bozek:2013uha}, but qualitatively 
the  same dependence of the average transverse momentum on rapidity is observed. In practice the LHC 
experiments cannot 
measure identified particles in a wide enough range of rapidities. The transverse momentum of charged particles 
as function of pseudorapidity from hydrodynamic calculations is shown in Fig. \ref{fig:2}. The change to the 
pseudorapidity variable causes a reduction of $\langle p_\perp(\eta)\rangle/\langle p_\perp(\eta=0)\rangle$ when going away from midrapidity, as compared to
$\langle p_\perp(y)\rangle/\langle p_\perp(y=0)\rangle$.
The effect is noticeable, but would not drive the value of  $\langle p_\perp(\eta\simeq 2)\rangle/\langle p_\perp(\eta=0)\rangle$ below 
one for the CGC case, so that 
experimentally the dependence of the average transverse momentum on pseudorapidity can be used to 
distinguish between the two scenarios.

As seen in Fig. \ref{fig:1}, going from midrapidity, $y=0$, towards the proton
fragmentation region we increase $\left\langle p_{\perp }\right\rangle $ in
the CGC owing to the increasing saturation momentum of the nucleus. On the
contrary, the $\left\langle p_{\perp }\right\rangle $ is decreasing in the
hydrodynamics picture owing to the decreasing number of particles. This is
the main result of our paper.

Finally we would like to make several remarks on the CGC expectations presented 
in this letter. The above results are reliable only for large $N_{\mathrm{part}}$ to
ensure that we have the separation of the two scales  $Q_{A}\gg Q_{p}$, 
which is implicitly assumed by applying the $k_t$-factorization formalism.
Our CGC results are based on quite general arguments and to obtain more precise predictions
a detailed model calculations should be performed, however this could be very challenging. For example, 
recently the NLO calculations in the CGC framework were performed in Ref.~\cite{Stasto:2013cha}. 
For the forward hadron production, these calculations demonstrated that the 
NLO corrections seem to be dominant at high transverse momentum. This implies that even higher 
order corrections should play an important role. 

\section{Conclusions}

In conclusion, we investigated the rapidity dependence of the average
transverse momentum of charged particles in proton-lead collisions at the LHC. We
noticed, based on the general arguments and simplified analytical
calculations, that, in the CGC, the transverse momentum is slightly increasing
with the increasing rapidity (going from $y=0$ towards the proton fragmentation region)
owing to the increasing saturation momentum of the nucleus. On the contrary, the 
$\left\langle p_{\perp }\right\rangle$ in the hydrodynamic framework is
decreasing owing to the decreasing number of particles. The collective expansion 
scenario cannot lead in a simple way to an increase of the average transverse momentum on the proton side.

We would like to point an interesting possibility, namely that $\langle p_\perp(y)\rangle$  decreases with $y$ 
around midrapidity, 
according to the collective expansion picture, but it starts to increase for larger $y$, in a region
where collectivity switches off and possibly saturation becomes dominant for the dynamics of the system.

\bigskip

\section*{Acknowledgments}
We thank A. Dumitru and L. McLerran for helpful discussions.  
P.B. is partly supported by the National Science Centre, 
Poland, grant DEC-2012/05/B/ST2/02528, and PL-Grid infrastructure.
A.B. is supported through the RIKEN-BNL Research Center.    

\bigskip 

%\bibliography{hydr}

\begin{thebibliography}{36}
\expandafter\ifx\csname natexlab\endcsname\relax\def\natexlab#1{#1}\fi
\providecommand{\bibinfo}[2]{#2}
\ifx\xfnm\relax \def\xfnm[#1]{\unskip,\space#1}\fi
%Type = Article
\bibitem[{Chatrchyan et~al.(2013)}]{CMS:2012qk}
\bibinfo{author}{S.~Chatrchyan}, et~al., \bibinfo{journal}{Phys. Lett.}
  \bibinfo{volume}{B718} (\bibinfo{year}{2013}) \bibinfo{pages}{795}.
%Type = Article
\bibitem[{Abelev et~al.(2013)}]{Abelev:2012ola}
\bibinfo{author}{B.~Abelev}, et~al., \bibinfo{journal}{Phys.Lett.}
  \bibinfo{volume}{B719} (\bibinfo{year}{2013}) \bibinfo{pages}{29}.
%Type = Article
\bibitem[{Aad et~al.(2013{\natexlab{a}})}]{Aad:2012gla}
\bibinfo{author}{G.~Aad}, et~al., \bibinfo{journal}{Phys.Rev.Lett.}
  \bibinfo{volume}{110} (\bibinfo{year}{2013}{\natexlab{a}})
  \bibinfo{pages}{182302}.
%Type = Article
\bibitem[{Aad et~al.(2013{\natexlab{b}})}]{Aad:2013fja}
\bibinfo{author}{G.~Aad}, et~al., \bibinfo{journal}{arXiv: 1303.2084 [hep-ex]}  (\bibinfo{year}{2013}{\natexlab{b}}).
%Type = Article
\bibitem[{Chatrchyan et~al.(2013)}]{Chatrchyan:2013nka}
\bibinfo{author}{S.~Chatrchyan}, et~al., \bibinfo{journal}{Phys.Lett.}
  \bibinfo{volume}{B724} (\bibinfo{year}{2013}) \bibinfo{pages}{213}.
%Type = Article
\bibitem[{Abelev et~al.(2013)}]{ABELEV:2013wsa}
\bibinfo{author}{B.~Abelev}, et~al., \bibinfo{journal}{arXiv: 1307.3237 [nucl-ex]}  (\bibinfo{year}{2013}).
%Type = Article
\bibitem[{Dusling et~al.(2010)Dusling, Gelis, Lappi, and
  Venugopalan}]{Dusling:2009ni}
\bibinfo{author}{K.~Dusling}, \bibinfo{author}{F.~Gelis},
  \bibinfo{author}{T.~Lappi}, \bibinfo{author}{R.~Venugopalan},
  \bibinfo{journal}{Nucl. Phys.} \bibinfo{volume}{A836} (\bibinfo{year}{2010})
  \bibinfo{pages}{159}.
%Type = Article
\bibitem[{Dusling and Venugopalan(2013)}]{Dusling:2013oia}
\bibinfo{author}{K.~Dusling}, \bibinfo{author}{R.~Venugopalan},
  \bibinfo{journal}{Phys. Rev.} \bibinfo{volume}{D87} (\bibinfo{year}{2013})
  \bibinfo{pages}{094034}.
%Type = Article
\bibitem[{Dusling and Venugopalan(2012)}]{Dusling:2012iga}
\bibinfo{author}{K.~Dusling}, \bibinfo{author}{R.~Venugopalan},
  \bibinfo{journal}{Phys.Rev.Lett.} \bibinfo{volume}{108}
  (\bibinfo{year}{2012}) \bibinfo{pages}{262001}.
%Type = Article
\bibitem[{Bo\.zek(2012)}]{Bozek:2011if}
\bibinfo{author}{P.~Bo\.zek}, \bibinfo{journal}{Phys. Rev.}
  \bibinfo{volume}{C85} (\bibinfo{year}{2012}) \bibinfo{pages}{014911}.
%Type = Article
\bibitem[{Bo\.zek and Broniowski(2013)}]{Bozek:2013uha}
\bibinfo{author}{P.~Bo\.zek}, \bibinfo{author}{W.~Broniowski},
  \bibinfo{journal}{Phys. Rev.} \bibinfo{volume}{C88} (\bibinfo{year}{2013})
  \bibinfo{pages}{014903}.
%Type = Article
\bibitem[{Bzdak et~al.(2013)Bzdak, Schenke, Tribedy, and
  Venugopalan}]{Bzdak:2013zma}
\bibinfo{author}{A.~Bzdak}, \bibinfo{author}{B.~Schenke},
  \bibinfo{author}{P.~Tribedy}, \bibinfo{author}{R.~Venugopalan},
  \bibinfo{journal}{Phys. Rev.} \bibinfo{volume}{C87} (\bibinfo{year}{2013})
  \bibinfo{pages}{064906}.
%Type = Article
\bibitem[{Qin and M\"uller(2013)}]{Qin:2013bha}
\bibinfo{author}{G.-Y. Qin}, \bibinfo{author}{B.~M\"uller}, \bibinfo{journal}{arXiv: 1306.3439 [nucl-th]}
  (\bibinfo{year}{2013}).
%Type = Article
\bibitem[{Werner et~al.(2013)Werner, Bleicher, Guiot, Karpenko, and
  Pierog}]{Werner:2013ipa}
\bibinfo{author}{K.~Werner}, \bibinfo{author}{M.~Bleicher},
  \bibinfo{author}{B.~Guiot}, \bibinfo{author}{I.~Karpenko},
  \bibinfo{author}{T.~Pierog}, \bibinfo{journal}{arXiv: 1307.4379 [nucl-th]}  (\bibinfo{year}{2013}).
%Type = Article
\bibitem{Bozek:2013ska} 
  P.~Bozek, W.~Broniowski and G.~Torrieri,
  %``Mass hierarchy in identified particle distributions in proton-lead collisions,''
  arXiv:1307.5060 [nucl-th] (2013).
  %%CITATION = ARXIV:1307.5060;%%
  %3 citations counted in INSPIRE as of 23 Sep 2013
%Type = Article
\bibitem[{Bo\.zek and Broniowski(2013)}]{Bozek:2013df}
\bibinfo{author}{P.~Bo\.zek}, \bibinfo{author}{W.~Broniowski},
  \bibinfo{journal}{Phys. Lett.} \bibinfo{volume}{B720} (\bibinfo{year}{2013})
  \bibinfo{pages}{250}.
%Type = Article
\bibitem[{Abelev et~al.(2013)}]{ALICE:2012mj}
\bibinfo{author}{B.~Abelev}, et~al., \bibinfo{journal}{Phys. Rev. Lett. 110,}
  \bibinfo{volume}{082302} (\bibinfo{year}{2013}) \bibinfo{pages}{082302}.
%Type = Article
\bibitem[{Chatrchyan et~al.(2013)}]{Chatrchyan:2013eya}
\bibinfo{author}{S.~Chatrchyan}, et~al., \bibinfo{journal}{arXiv: 1307.3442 [hep-ex]}   (\bibinfo{year}{2013}).
%Type = Article
\bibitem[{Abelev et~al.(2013)}]{Abelev:2013bla}
\bibinfo{author}{B.~B. Abelev}, et~al., \bibinfo{journal}{arXiv: 1307.1094 [nucl-ex]}  (\bibinfo{year}{2013}).
%Type = Article
\bibitem[{Bzdak and Skokov(2013)}]{Bzdak:2013lva}
\bibinfo{author}{A.~Bzdak}, \bibinfo{author}{V.~Skokov}, \bibinfo{journal}{arXiv: 1306.5442 [nucl-th]}
  (\bibinfo{year}{2013}).
%Type = Article
\bibitem[{Ortiz et~al.(2013)Ortiz, Christiansen, Cuautle, Maldonado, and
  Paic}]{Ortiz:2013yxa}
\bibinfo{author}{A.~Ortiz}, \bibinfo{author}{P.~Christiansen},
  \bibinfo{author}{E.~Cuautle}, \bibinfo{author}{I.~Maldonado},
  \bibinfo{author}{G.~Paic}, \bibinfo{journal}{Phys.Rev.Lett.}
  \bibinfo{volume}{111} (\bibinfo{year}{2013}) \bibinfo{pages}{042001}.
%Type = Article
\bibitem[{McLerran et~al.(2013)McLerran, Praszalowicz, and
  Schenke}]{McLerran:2013oju}
\bibinfo{author}{L.~McLerran}, \bibinfo{author}{M.~Praszalowicz},
  \bibinfo{author}{B.~Schenke}, \bibinfo{journal}{arXiv: 1308.2350 [hep-ph]}  (\bibinfo{year}{2013}). 
%Type = Article
\bibitem[{Bzdak and Skokov(2013)}]{Bzdak:2013zla}
\bibinfo{author}{A.~Bzdak}, \bibinfo{author}{V.~Skokov}, \bibinfo{journal}{arXiv: 1307.6168 [hep-ph]}  
  (\bibinfo{year}{2013}).
%Type = Article
\bibitem[{Kovchegov and Tuchin(2002)}]{Kovchegov:2001sc}
\bibinfo{author}{Y.~V. Kovchegov}, \bibinfo{author}{K.~Tuchin},
  \bibinfo{journal}{Phys.Rev.} \bibinfo{volume}{D65} (\bibinfo{year}{2002})
  \bibinfo{pages}{074026}.
%Type = Article
\bibitem[{Dumitru and McLerran(2002)}]{Dumitru:2001ux}
\bibinfo{author}{A.~Dumitru}, \bibinfo{author}{L.~D. McLerran},
  \bibinfo{journal}{Nucl.Phys.} \bibinfo{volume}{A700} (\bibinfo{year}{2002})
  \bibinfo{pages}{492}.
%Type = Article
\bibitem[{Kharzeev et~al.(2005)Kharzeev, Levin, and Nardi}]{Kharzeev:2004if}
\bibinfo{author}{D.~Kharzeev}, \bibinfo{author}{E.~Levin},
  \bibinfo{author}{M.~Nardi}, \bibinfo{journal}{Nucl.Phys.}
  \bibinfo{volume}{A747} (\bibinfo{year}{2005}) \bibinfo{pages}{609}.
%Type = Article
\bibitem[{Khoze et~al.(2004)Khoze, Martin, Ryskin, and Stirling}]{Khoze:2004hx}
\bibinfo{author}{V.~Khoze}, \bibinfo{author}{A.~Martin},
  \bibinfo{author}{M.~Ryskin}, \bibinfo{author}{W.~Stirling},
  \bibinfo{journal}{Phys.Rev.} \bibinfo{volume}{D70} (\bibinfo{year}{2004})
  \bibinfo{pages}{074013}.
%Type = Article
\bibitem[{Dumitru et~al.(2012)Dumitru, Kharzeev, Levin, and
  Nara}]{Dumitru:2011wq}
\bibinfo{author}{A.~Dumitru}, \bibinfo{author}{D.~E. Kharzeev},
  \bibinfo{author}{E.~M. Levin}, \bibinfo{author}{Y.~Nara},
  \bibinfo{journal}{Phys.Rev.} \bibinfo{volume}{C85} (\bibinfo{year}{2012})
  \bibinfo{pages}{044920}.
%Type = Article
\bibitem[{Abelev et~al.(2013)}]{ALICE:2012xs}
\bibinfo{author}{B.~Abelev}, et~al., \bibinfo{journal}{Phys. Rev. Lett. 110,}
  \bibinfo{volume}{032301} (\bibinfo{year}{2013}) \bibinfo{pages}{032301}.
%Type = Article
\bibitem[{Aad et~al.(2013)}]{ATLAS-CONF-2013-096}
\bibinfo{author}{G.~Aad}, et~al.,   \bibinfo{journal}{ATLAS-CONF-2013-096}  (\bibinfo{year}{2013}).
%Type = Article
\bibitem[{Ollitrault(1991)}]{Ollitrault:1991xx}
\bibinfo{author}{J.-Y. Ollitrault}, \bibinfo{journal}{Phys.Lett.}
  \bibinfo{volume}{B273} (\bibinfo{year}{1991}) \bibinfo{pages}{32}.
%Type = Article
\bibitem[{Bo\.zek(2012)}]{Bozek:2011ua}
\bibinfo{author}{P.~Bo\.zek}, \bibinfo{journal}{Phys. Rev.}
  \bibinfo{volume}{C85} (\bibinfo{year}{2012}) \bibinfo{pages}{034901}.
%Type = Article
\bibitem[{Back et~al.(2003)}]{Back:2003ns}
\bibinfo{author}{B.~B. Back}, et~al., \bibinfo{journal}{Phys. Rev. Lett.}
  \bibinfo{volume}{91} (\bibinfo{year}{2003}) \bibinfo{pages}{072302}.
%Type = Article
\bibitem[{Bia\l{}as and Czy\.z(2005)}]{Bialas:2004su}
\bibinfo{author}{A.~Bia\l{}as}, \bibinfo{author}{W.~Czy\.z},
  \bibinfo{journal}{Acta Phys. Polon.} \bibinfo{volume}{B36}
  (\bibinfo{year}{2005}) \bibinfo{pages}{905}. 
%Type = Article
\bibitem[{Bia\l{}as and Bzdak(2008)}]{Bialas:2007eg}
\bibinfo{author}{A.~Bia\l{}as}, \bibinfo{author}{A.~Bzdak},
  \bibinfo{journal}{Phys. Rev.} \bibinfo{volume}{C77}
  (\bibinfo{year}{2008}) \bibinfo{pages}{034908}.     
%Type = Article
\bibitem[{Stasto et~al.(2013)Stasto, Xiao, and Zaslavsky}]{Stasto:2013cha}
\bibinfo{author}{A.~M. Stasto}, \bibinfo{author}{B.-W. Xiao},
  \bibinfo{author}{D.~Zaslavsky}, \bibinfo{journal}{arXiv: 1307.4057 [hep-ph]}   (\bibinfo{year}{2013}).

\end{thebibliography}

\end{document}